\documentstyle[12pt,aasms4]{article}

\newcommand{\wone}{\ifmmode{\nu_{\rm 1H}}\else{$\nu_{\rm 1H}$}\fi}
\newcommand{\wtwo}{\ifmmode{\nu_{\rm 2H}}\else{$\nu_{\rm 2H}$}\fi}
\newcommand{\ord}{\ifmmode{{\cal O}}\else{{\cal O}}\fi}
\newcommand{\PL}{\ifmmode{P\hbox{--}L}\else{$P\hbox{--}L$}\fi}
\newcommand{\ML}{\ifmmode{M\hbox{--}L}\else{$M\hbox{--}L$}\fi}
\newcommand{\ie}{{\it i.e.}}
\newcommand{\eg}{{\it e.g.}}
\newcommand{\etal}{{\it et al.}}

\lefthead{Alcock \etal}
\righthead{2H Pulsation Mode from SMC 1H/2H Beat Cepheids}

\begin{document}

%
%
\title{The MACHO Project SMC Variable Star Inventory. I. The 
Second-overtone Mode of Cepheid Pulsation From First/Second Overtone (1H/2H)
Beat Cepheids}

\author{C.~Alcock\altaffilmark{1,2}, 
      R.A.~Allsman\altaffilmark{3},
        D.~Alves\altaffilmark{1,4},
      T.S.~Axelrod\altaffilmark{5},
      A.C.~Becker\altaffilmark{6},
      D.P.~Bennett\altaffilmark{1,2,7},
      K.H.~Cook\altaffilmark{1,2},
      K.C.~Freeman\altaffilmark{5},
        K.~Griest\altaffilmark{2,8},
      M.J.~Lehner\altaffilmark{2,8},
      S.L.~Marshall\altaffilmark{1,2},
      B.A.~Peterson\altaffilmark{5},
      P.J.~Quinn\altaffilmark{10},
      A.W.~Rodgers\altaffilmark{5},
      A.~Rorabeck\altaffilmark{12},
        W.~Sutherland\altaffilmark{11},
        A.~Tomaney\altaffilmark{6},
        T.~Vandehei\altaffilmark{2,8}\\
      {\bf (The MACHO Collaboration) }
      }

\altaffiltext{1}{Lawrence Livermore National Laboratory, Livermore, CA 94550
        E-mail: {\tt alcock, alves, dminniti, kcook, stuart@igpp.llnl.gov}}
 
\altaffiltext{2}{Center for Particle Astrophysics,
        University of California, Berkeley, CA 94720}
 
\altaffiltext{3}{Supercomputing Facility, Australian National University,
        Canberra, ACT 0200, Australia
        E-mail: {\tt robyn@macho.anu.edu.au}}
 
\altaffiltext{4}{Department of Physics, University of California,
        Davis, CA 95616 }
 
\altaffiltext{5}{Mt.~Stromlo and Siding Spring Observatories, Australian
        National University, Weston Creek, ACT 2611, Australia
        E-mail: {\tt tsa, kcf,peterson, alex@mso.anu.edu.au}}
 
\altaffiltext{6}{Departments of Astronomy and Physics,
        University of Washington, Seattle, WA 98195
        E-mail: {\tt austin, becker, mrp@astro.washington.edu}}

\altaffiltext{7}{Physics Department, University of Notre Dame, Notre 
        Dame, IN 46556
        E-mail: {\tt bennett@bustard.phys.nd.edu}}
 
\altaffiltext{8}{Department of Physics, University of California,
        San Diego, La Jolla, CA 92093
        E-mail: {\tt kgriest, tvandehei, mlehner@ucsd.edu }}
 
\altaffiltext{9}{Department of Physics, University of California,
        Santa Barbara, CA 93106}
 
\altaffiltext{10}{European Southern Observatory, Karl-Schwarzchild Str. 2,
        D-85748, Garching, Germany
        E-mail: {\tt pjq@eso.org}}

\altaffiltext{11}{Department of Physics, University of Oxford, Oxford OX1
        3RH, U.K.
        E-mail: {\tt w.sutherland@physics.ox.ac.uk}}
 
\altaffiltext{12}{Dept. of Physics \& Astronomy, McMaster University,
        Hamilton, Ontario, L8S 4M1 Canada
        E-mail: {\tt welch, rorabec@physics.mcmaster.ca}}

%
%
\setcounter{footnote}{0}

%
%
\begin{abstract}

We report the discovery of 20 1H/2H and 7 F/1H beat Cepheids in the SMC by
the MACHO Project.  We utilize the 20 1H/2H stars to determine lightcurve
shape for the SMC second-overtone (2H) mode of Cepheid pulsation.  We
predict, similar to the findings of Alcock \etal\ [\markcite{beat2}1997,
ApJ, submitted], that 2H Cepheids will have nearly or purely sinusoidal
light variations; that the \PL\ relation for 2H Cepheids will not be
distinguishable from the \PL\ relation for 1H Cepheids within photometric
accuracy; and that 2H stars may be discernable from F and 1H stars using
the amplitude-period diagram and Fourier parameter progressions for
periods $P\lesssim 0.7$ days, our current sample 2H period limit. 

\end{abstract}

%
%

\section{Introduction}

The second-overtone (2H) mode of Cepheid pulsation has been predicted to
exist theoretically since Stobie (\markcite{Stobiea}1969a,
\markcite{Stobieb}1969b)'s pioneering investigations. Yet, since then, we
have found only scant evidence for 2H mode excitation in our Galaxy.  CO
Aur was recognized as a first-overtone/second-overtone (1H/2H) beat
Cepheid by Mantegazza \markcite{mante}(1983), and later confirmed as such
by various studies (\eg, Antonello \& Mantegazza \markcite{antmant}1984;
Babel \& Burki \markcite{BabBurk}1987).  On the other hand, HR 7308 is a
proposed singly-periodic 2H Cepheid whose modal status remains uncertain,
despite many investigations (Burki \etal\ \markcite{BSAF}1986; Fabregat,
Suso, \& Reglero \markcite{Fab}1990; Simon \markcite{Simon85}1985; Bersier
\markcite{Ber96}1996; Bersier \& Burki \markcite{BerBur}1996).  This
paucity of Galactic 2H Cepheids is not unexpected.  From a theoretical
standpoint, Galactic 2H Cepheids should have low masses and luminosities
(Chiosi, Wood \& Capitano \markcite{CWC}1993).  As well, they are expected
to be the shortest-period Cepheids at a given luminosity (Chiosi \etal\
\markcite{CWC}1993), so that they should appear in greater frequency in
lower metallicity environments than our own (see \eg, the period frequency
distributions of Cepheids in Lipunova \markcite{Lip}1992). 
Observationally, in our own galaxy, CO Aur's semi-amplitude of pulsation
for its 2H mode is only $0.043\pm0.002$ mag (Pardo \& Poretti
\markcite{Pardo}1996)---so that, even if we observe these faint stars, we
might not detect their variability. 

The advent of large-scale astronomical surveys has improved our chances of
observing 2H Cepheids.  As by-products of gravitational microlensing
searches in the Galactic bulge and Magellanic Clouds, the MACHO and EROS
Collaborations have found 45 1H/2H and at least 37 F/1H beat Cepheids in
the LMC (Alcock \etal\ \markcite{beat1}1995, \markcite{beat2}1997; 
Beaulieu \etal\ \markcite{EROS}1997), and 27 1H/2H and 10 F/1H beats
(counting this work and Beaulieu \etal\ \markcite{EROS}1997) in the SMC to
date.  Concurrent analyses of these, and other findings, has allowed
investigations of the 2H mode of Cepheid pulsation. Pardo \& Poretti
\markcite{Pardo}(1996) re-analyzed the composite lightcurve of CO Aur, the
sole 1H/2H beat Cepheid in the Galaxy, and noted that its 2H mode appeared
as a purely sinusoidal light variation. Alcock \etal\
\markcite{beat2}(1997) analyzed 45 first-overtone/second-overtone beat
Cepheids in the LMC, showing (1) that the 2H mode resulted in sinusoidal,
or nearly sinusoidal ($0\leq R_{21}\lesssim 0.2$)
lightcurves\footnote{$R_{k1}=V_k/V_1$ is the relative amplitude of the
first harmonic and `base' frequency model amplitudes in a truncated
Fourier series $V(t) = V_0 + \sum_{k = 1}^\ord V_k\cos(2\pi k\nu
t+\phi_k)$, while the phase difference $\phi_{k1} = \phi_k-k\phi_1$.  For
beat Cepheids, $R_{21}$ and $\phi_{21}$ are calculated for each mode of
pulsation.}; (2) that LMC 2H Cepheids could be distinguished from LMC 1H
and F Cepheids in Fourier space for $P\lesssim 1.4$ days; (3) that 2H
Cepheids should overlap the short-period edge of the 1H \PL\ sequence; 
and (4) that the location of 2H pulsators in the $\log L$--$\log T_{\rm
eff}$ plane depended significantly on the adopted \ML\ relation, and would
have to come from observation.  Finally, Antonello \& Kanbur
\markcite{antkan}(1997) have investigated the 2H mode of Cepheid pulsation
by non-linear pulsation models appropriate to the LMC ($Z\approx 0.01$). 
They confirmed that 2H Cepheids should be more numerous for lower
metallicities, and produced theoretical $R_{21}$--$P$ sequences which
agreed qualitatively with the sequences for LMC 1H/2H beat Cepheids in
Welch \etal\ \markcite{Welch}(1997). They also predicted a resonance of
the $R_{21}$--$P$ and $\phi_{21}$--$P$ sequences near $P=1$ day. 

With the recent reduction of SMC photometry by the MACHO project, we are
in a position to add to our knowledge of the 2H mode. We report the
discovery of 20 1H/2H beat Cepheids in the SMC (distinct from the stars in
Beaulieu \etal\ \markcite{EROS}1997), and their implications for the 2H
mode of Cepheid pulsation.  We compare our findings to the 2H mode
characterizations in the LMC and Galaxy to date, and provide guidance on
how to discern 2H from F and 1H Cepheids. 

\section{Observations and Analysis}

We refer the reader to Alcock \etal\ \markcite{beat1}(1995) for a
description of our two-bandpass photometry (the MACHO $V$ and $R$ bands)
and beat Cepheid identification process.  The beat Cepheids reported in
this paper were selected by our alert system software and not by a full
analysis run. Therefore, the total number of beat Cepheids in these fields
is likely to be 4-5 $\times$ greater than reported here. To be identified
as an alert, a star must be 7 sigma brighter than the template and have
increased in brightness by at least 0.35 mag.  SMC observations of these
Cepheids span 3 years;  lightcurves consist of anywhere from 163--1306
observations, which are free of possible cosmic ray events, bad or missing
pixels, or data suffering from poor image quality.  This paper utilizes
MACHO $V$-band photometry for all results. 

We subjected each star to our coding of the CLEANest algorithm (Foster
\markcite{F95}1995, \markcite{F96a}1996a, \markcite{F96b}1996b) for joint
frequency analysis and lightcurve modelling.  This method avoids having to
choose a truncated Fourier series order {\it a priori}, as discussed in
Pardo \& Poretti \markcite{Pardo}(1997) and Alcock \etal\
\markcite{beat2}(1997).  Briefly, CLEANest uses the date-compensated
discrete Fourier transform (DCDFT) of Ferraz-Mello \markcite{FM}(1981) on
a time series to produce a power spectrum for test frequencies from
$\nu_{\rm res} = (2T_{\rm span})^{-1}$ to $\nu_{\rm max} = (2\min(\Delta
t))^{-1}$ in steps of $\nu_{\rm res}$ (the frequency resolution), where
$T_{\rm span}$ is the total timespan of the observations for a star, and
$\Delta t$ the time separation between successive observations. If any of
the frequencies in the power spectrum are adopted as significant, they are
modeled by $\cos(2\pi\nu t)$ and $\sin(2\pi\nu t)$ terms (plus a constant)
as in Foster \markcite{F95}(1995).  The resultant model is subtracted from
the data, these {\it residuals} are subjected to another DCDFT, and the
process is iterated until no significant frequencies remain.  Each time a
DCDFT of the data or residuals has been performed, CLEANest seeks to find
the $n$-tuple of frequencies which gives the best description of the data.
Operationally, frequency space is searched for frequencies in the
neighbourhood of the currently adopted ones for a maximum of Foster
(\markcite{F96a}1996a, \markcite{F96b}1996b)'s model amplitude. 

In all cases, the 1H mode frequency, \wone, appeared as the peak frequency
in the first power spectrum, generally followed either by 2\wone\ or
\wtwo.  We confirmed the identity of \wone\ and \wtwo\ by requiring
$\wone/\wtwo \sim 0.805$, as found in Alcock \etal\
\markcite{beat1}(1995).  After these frequencies were discovered, we
adopted a frequency as significant if it was a linear combination of
\wone\ and \wtwo; if it appeared as one of the 20 most powerful
frequencies in a residual spectrum; {\it and} if it was reasonable (\ie,
we would not have modeled a frequency that seemed to be $2\wone + \wtwo$
if we had not previously detected 2\wone\ or \wtwo\ in our analysis), as
discussed in Alcock \etal\ \markcite{beat2}(1997).  For some stars we had
to adopt a frequency close to 1.003 day$^{-1}$ (\ie, the frequency
corresponding to one sidereal day) in the modeling process, because of the
scheduling of observations.  When no remaining significant frequencies
could be detected, we discontinued modeling with CLEANest, and subjected
the fits to the Marquardt algorithm\footnote{The Marquardt algorithm is a
$\chi^2$ minimization method which pragmatically alternates between a
`steepest descent' (or gradient-search) algorithm when $\chi^2$ changes
rapidly near a given set of model parameters, and a first-order model
expansion when $\chi^2$ changes little near a set of model parameters (see
\eg, Bevington \& Robinson \markcite{Bev}1992).} for improvement, at no
point restricting our model frequencies to obey their expected relations
to \wone\ and \wtwo: \ie, all frequencies were varied by both CLEANest and
the Marquardt algorithm independently of \wone\ or \wtwo\ themselves. 
This was done as a check on the robustness and identity of a given
frequency. 

In Table \ref{pertable}, we present periods and period ratios for our 27
SMC beat Cepheids.  $P_{\rm L}$ is the `long' period, while $P_{\rm S}$ is
the `short' period, of pulsation.  For F/1H stars ($P_{\rm S}/P_{\rm L}
\sim 0.73$), $P_{\rm L} = P_{\rm F}$ and $P_{\rm S} = P_{\rm 1H}$; for
1H/2H stars ($P_{\rm S}/P_{\rm L} \sim 0.805$), $P_{\rm L} = P_{\rm 1H}$
and $P_{\rm S} = P_{\rm 2H}$.  Uncertainties in the last three digits of
periods and period ratios have been placed in parentheses; uncertainties
in the period ratios were obtained from the uncertainties in $P_{\rm L}$
and $P_{\rm S}$. 

\section{Results and Discussion}

\subsection{The Petersen Diagram}

We have plotted the Petersen diagram for all of our SMC beat Cepheids and
the 45 LMC 1H/2H beat Cepheids of Alcock \etal\ \markcite{beat2}(1997) as
Figure \ref{petersen}.  We note the LMC and SMC 1H/2H beat Cepheids have
essentially the same progression of period ratio $P_{\rm 2H}/P_{\rm 1H}$
versus $P_{\rm 1H}$ despite differences in host galaxy metallicity.  This
was borne out in the linear, non-adiabatic calculations of Morgan \& Welch
\markcite{MW}(1996), who predicted little or no noticeable shift in
$P_{\rm 2H}/P_{\rm 1H}$ from the LMC to SMC.  Observationally, this
similarity in period ratio was also noted by Beaulieu \etal\
\markcite{EROS}(1997) by comparing their 7 SMC 1H/2H beat Cepheids with
the 15 1H/2H beat Cepheids of Alcock \etal\ \markcite{beat1}(1995). 

\subsection{The Bailey Diagram}

The Bailey (or Period-amplitude) diagram for the 1H/2H beat Cepheids of
this paper and Alcock \etal\ \markcite{beat2}(1997) is presented as Figure
\ref{bailey}.  The amplitude $\Delta$ is the model semi-amplitude for the
base Fourier term of a mode of pulsation.  We see that, in general,
$\Delta_{\rm 2H}\lesssim 0.10$ mag, while $\Delta_{\rm 1H}\gtrsim 0.11$
mag, although this is not always the case: there is a star with
$\Delta_{\rm 2H} = 0.105$ mag in the SMC, and a star with $\Delta_{\rm 1H}
= 0.071$ mag in the LMC in the Figure.  Similar conclusions can be drawn
from Figure 3 of Beaulieu \etal\ \markcite{EROS}(1997), who found SMC 2H
mode amplitudes less than 0.08 mag, and 1H mode amplitudes of 0.12 mag for
similar periods. Clearly, the 2H mode results in a low pulsation
amplitude. 

\subsection{Fourier Parameter Sequences and 2H Mode Lightcurve Shape}

We detected the second harmonic of the 2H mode frequency, 2\wtwo, for 6 of
20 stars, but found this frequency remained stable (\ie, differed by more
than twice its formal uncertainty from twice \wone) in the Marquardt
improvement for only 2 of these 6 stars.  This is contrary to the LMC
beats analyzed by Alcock \etal\ \markcite{beat2}(1997): all 8 of 45 stars
that had 2\wtwo\ detected also had that frequency remain stable over
Marquardt improvement.  This may suggest 2\wtwo\ detections here are of a
more marginal nature than in the LMC, owing to the SMC's greater distance
modulus ($\mu_{\rm SMC}-\mu_{\rm LMC} \sim 0.5 - 0.7$ mag; Feast
\markcite{Feast88}1988, \markcite{Feast89}1989), and thus fainter stars,
although SMC exposures were twice as long as LMC exposures to compensate. 

No second (3\wtwo) or higher harmonics were detected in any of our sample
for the 2H mode, limiting us to $R_{21}$ and $\phi_{21}$ to describe its
lightcurve shape.  For the 1H mode, frequencies up to the third harmonic,
4\wone, were detected; the higher order $R_{31}-R_{41}$ and
$\phi_{31}-\phi_{41}$ for the 1H mode will be presented in a future paper. 

Fourier parameters for those stars with detected, stable harmonics in our
CLEANest--Marquardt scheme are presented in Figure \ref{Four21}. 
According to this Figure, 2 of 20 1H/2H beat Cepheids have nearly
sinusoidal lightcurves for their 2H modes, while the remaining 18 have 2H
modes that result in purely sinusoidal light variations: \ie, $R_{21}=0$,
as shown in the Figure. 

The stable 2\wtwo\ frequencies in the LMC 1H/2H beat Cepheids of Alcock
\etal\ \markcite{beat2}(1997) were what prompted us to draw out 2H mode
information from the LMC sample.  We would like to gather as much
information as possible on the 2H mode here as well. To circumvent
stability concerns with 2\wtwo\ frequencies, we have adopted a first
harmonic term for each 1H/2H star's model, and fit it to our data while
holding its frequency to its expected value of 2\wtwo. No other
frequencies were held to their expected identities, as they retained their
relationships to \wone\ and \wtwo\ throughout. We display the resulting
Fourier parameter sequences in Figure \ref{Four21-2} for those stars with
$\sigma_{R_{21}} < 0.05$, which omitted 3 of 20 stars.  The conclusions
from either Figure \ref{Four21} or \ref{Four21-2} are the same: in
general, the 2H mode is more sinusoidal than the 1H mode (from $R_{21}$). 
The scatter in $\phi_{21}$ prevents further comment. 

\section{Discerning 2H Cepheids}

We do not yet have the luxury of a large SMC sample of Fourier-decomposed
Cepheids on the same photometric system---or transformations to standard
systems for the SMC---to compare our $R_{21}$--$P$ and $\phi_{21}$--$P$
diagrams against.  Our sample of beat Cepheids also constitutes a
`first-order' search through our developing SMC photometry database, and
so cannot be claimed as complete.  This is the first large sample of SMC
beat Cepheids, however, so we should use them to make some statements
about the SMC 2H mode. 

Firstly, we note Figure \ref{petersen} and Table \ref{pertable} show the
1H and 2H \PL\ relations will be separated by less than $\log(P_{\rm
2H}/P_{\rm 1H}) = \log 0.80 \simeq -0.10$ in the SMC.  Alcock \etal\
\markcite{beat1}(1995) noted that this separation could well vanish due to
observational uncertainties, overlapping the 2H and 1H \PL\ sequences. 
This suggests separation of SMC 1H and 2H mode Cepheids based on \PL\
position alone is not feasible. 

Alcock \etal\ \markcite{beat2}(1997) used the analytical fits to linear
nonadiabatic pulsation calculations of Chiosi \etal\ \markcite{CWC}(1993)
to see how the boundaries of the instability strip (IS) and \PL\ relations
in the LMC were affected by the choice of \ML\ relation.  They chose to
use the evolutionary \ML\ relation from mild core-overshoot models of
Chiosi \etal\ \markcite{CWC}(1993) and the empirical \ML\ relation of
Simon \markcite{simon90}(1990), and found that (1) relative positions for
the F, 1H and 2H modes in the IS are markedly affected by the adopted \ML\
relation, so that the positions of 2H pulsators in the IS (or CMD) will
have to come from observation; while (2) relative \PL\ positions for the
F, 1H and 2H modes remain, from shortest to longest period at a given
luminosity: 2H, then 1H and F mode pulsators.  These same conclusions hold
for the SMC, using the same \ML\ relations and compositions of $Y=0.30$,
$Z=0.004$: we reproduce the same trends and general positions of mode
boundaries as Figures 1-4 of Alcock \etal\ \markcite{beat2}(1997). 

Fourier parameters provide better distinction than \PL\ or IS position
between the 1H and 2H modes in the SMC.  Figure \ref{Four21} shows, as
already found for the 2H mode of Cepheid pulsation in the LMC, that {\rm
2H Cepheids will have nearly, or purely, sinusoidal light variations},
which should allow them to be discerned from 1H and F mode stars.  The
period-amplitude diagram (Figure \ref{bailey}) should provide further
support for a 1H-2H mode distinction.  This, of course, ignores sources of
contamination, such as foreground W Ursa Majoris stars (\eg, Kaluzny,
Thompson \& Krzeminski \markcite{WUMa}1997), which can have nearly
sinusoidal light variations and yet may occupy the same CMD and \PL\
regions as SMC or LMC Cepheids.  Given a {\it bona fide} Cepheid, however,
we should---with some certainty---be able to discern in which mode it
pulsates from the information available to us. 

\acknowledgements

We are grateful for the skilled support given our project by the technical
staff at Mt. Stromlo Observatory (MSO). Work performed at Lawrence
Livermore National Laboratory (LLNL) is supported by the Department of
Energy (DOE) under contract W7405-ENG-48. Work performed by the Center for
Particle Astrophysics (CfPA) on the University of California campuses is
supported in part by the Office of Science and Technology Centers of the
National Science Foundation (NSF) under cooperative agreement AST-8809616. 
Work performed at MSO is supported by the Bilateral Science and Technology
Program of the Australian Department of Industry, Technology and Regional
Development. KG acknowledges a DOE OJI grant, and the support of the Sloan
Foundation. DLW and AJR were supported, in part, by a Research Grant from
the Natural Sciences and Engineering Research Council of Canada (NSERC)
during this work.  AJR was also supported, in part, by an NSERC
Postgraduate scholarship (PGS A).  This work comprises part of his M.Sc. 
thesis.

\newpage
\figcaption[petersen.eps]{The Petersen diagram for all 27 MACHO Project 
SMC beat Cepheids, as well as the 45 1H/2H LMC beat Cepheids of Alcock 
\etal\ (1997).  The upper sequence is $P_{\rm 2H}/P_{\rm 
1H}$ vs. $P_{\rm 1H}$, while the lower sequence is $P_{\rm 1H}/P_{\rm 
F}$ vs. $P_{\rm F}$.\label{petersen}}

\figcaption[bailey.eps]{Amplitude-period diagram for 1H/2H beat Cepheids 
in the Magellanic Clouds from MACHO Project data.  $\Delta$ is the model
semi-amplitude of the base Fourier term for each of the 1H (open points) 
and 2H (filled points) modes.\label{bailey}}

\figcaption[Four21.eps]{Fourier parameters for the 1H, 2H, and F modes of 
pulsation from 27 SMC beat Cepheids.\label{Four21}.}

\figcaption[Four21-2.eps]{Fourier parameters for the 20 SMC 1H/2H beat 
Cepheids when fitted by a first harmonic term for the 2H mode.  Stars with
$\sigma_{R_{21}} < 0.05$ are not shown.\label{Four21-2}}

\begin{table}
\dummytable\label{pertable}
\end{table}

\end{document}